\documentclass[aps,superscriptaddress,prl,reprint,amsmath,floatfix]{revtex4-1}
\usepackage{graphicx}
\usepackage{array}
\usepackage{booktabs}
\usepackage[colorlinks, linkcolor=blue]{hyperref} 
\usepackage{bm}
\usepackage{color}
\usepackage{subfigure}
\usepackage{soul}
\usepackage{amsmath}
\usepackage{amsfonts}
\usepackage{amssymb}
\usepackage{lipsum}

\newcommand{\DpRP}{\ensuremath{\delta_\textsc{rp}}}
\newcommand{\DpB}{\ensuremath{\delta_\textsc{j}}}

\newcommand{\XL}{\ensuremath{\chi_\textsc{l}}}

\definecolor{darkBlue}{rgb}{0,0,0.6}
\definecolor{darkRed}{rgb}{0.5,0,0}
\definecolor{darkGreen}{rgb}{0,0.5,0}

\begin{document}
\title{Universal scaling for disordered viscoelastic matter near the onset of rigidity}
\author{Danilo B. Liarte}
\email{danilo.liarte@ictp-saifr.org}
\affiliation{ICTP South American Institute for Fundamental Research, S\~ao Paulo, SP, Brazil}
\affiliation{Institute of Theoretical Physics, S\~ao Paulo State University, S\~ao Paulo, SP, Brazil}
\affiliation{Department of Physics, Cornell University, Ithaca, NY, 14853, USA}
\author{Stephen J. Thornton}
\author{Eric Schwen}
\author{Itai Cohen}
\author{Debanjan Chowdhury}
\author{James P. Sethna}
\affiliation{Department of Physics, Cornell University, Ithaca, NY, 14853, USA}
\date{\today}
	
\begin{abstract}
The onset of rigidity in interacting liquids, as they undergo a transition to a disordered solid, is associated with a rearrangement of the low-frequency vibrational spectrum. In this letter, we derive scaling forms for the singular dynamical response of disordered
viscoelastic networks near both jamming and rigidity percolation.
Using effective-medium theory, we extract
critical exponents, invariant scaling combinations and analytical formulas for 
universal scaling functions near these transitions.
Our scaling forms describe the behavior in space and time near the various onsets of rigidity, for rigid and floppy phases and the crossover region, including diverging length and time scales at the transitions.
\end{abstract}

\maketitle

\emph{Jamming}~\cite{LiuNag2010} and \emph{Rigid Percolation} (RP)~\cite{Thorpe1983} provide suitable frameworks to characterize the fascinating invariant scaling behavior exhibited by several classes of disordered viscoelastic materials near the onset of rigidity~\cite{SethnaZap2017}.
Both are often described by elastic networks near the Maxwell limit of mechanical stability~\cite{Maxwell1864}, and represent transitions from a rigid phase to a floppy one when the average coordination number $z$ falls below the isostatic value $z_c$.
RP appears in network glasses~\cite{Thorpe2002}, fiber networks~\cite{Picu2011,BroederszMac2011} and soft colloidal gels~\cite{ZhangMao2019}, and is described in terms of networks in which bonds are randomly removed; the bulk modulus vanishes%
~\footnote{
Different types of lattices do not {\em appear} to have the same universal RP behavior.
For instance, isotropic periodic Maxwell lattices (in which $z=z_c=2D$ where $D$
is the dimension) can have $B, G>0$ (as in the kagome lattice), where $G$ is the
shear modulus, or $B=0$ and $G>0$ (as in the twisted-kagome lattice, see e.g.~\cite{LiarteLub2020}), which suggests that these lattices do not belong to the same RP universality class.
However, if these lattices have extra bonds so that $z>2D$, arbitrary protocols to randomly dilute these networks without specifically targeting particular bonds will lead to a continuous transition for both $B$ and $G$.
}
at the transition~\cite{FengGar1985,MaoLub2011,LiarteLub2016}.
Jamming is also a ubiquitous phenomenon arising in systems ranging from amorphous solids and glasses~\cite{LiuWya2011} to cell tissues~\cite{BiMan2016} and deep learning~\cite{BahriGan2020}. Jamming is commonly described in terms of sphere packings that possess a finite bulk modulus $B>0$ at the transition. Recently, it was shown that jamming can be described as a multi-critical point that terminates a line of continuous transitions associated with rigidity percolation and that there is a deep connection between the universal scaling forms for both transitions~\cite{LiarteLub2019}. Determining explicit formulas for the susceptibilities and space-time correlations has been challenging, however, since there is a scarcity both of comprehensive numerical data and of analytic models for these transitions (with the exception of jamming in high dimensions~\cite{KurchanZam2012,KurchanZam2013,CharbonneauZam2014}). Here, we leverage the analytically-tractable effective-medium theory (EMT) of Ref.~\cite{LiarteLub2019} to fill this gap and extract explicit equations for these universal forms.

\begin{figure}[!ht]
\begin{center}
\includegraphics[width=\linewidth]{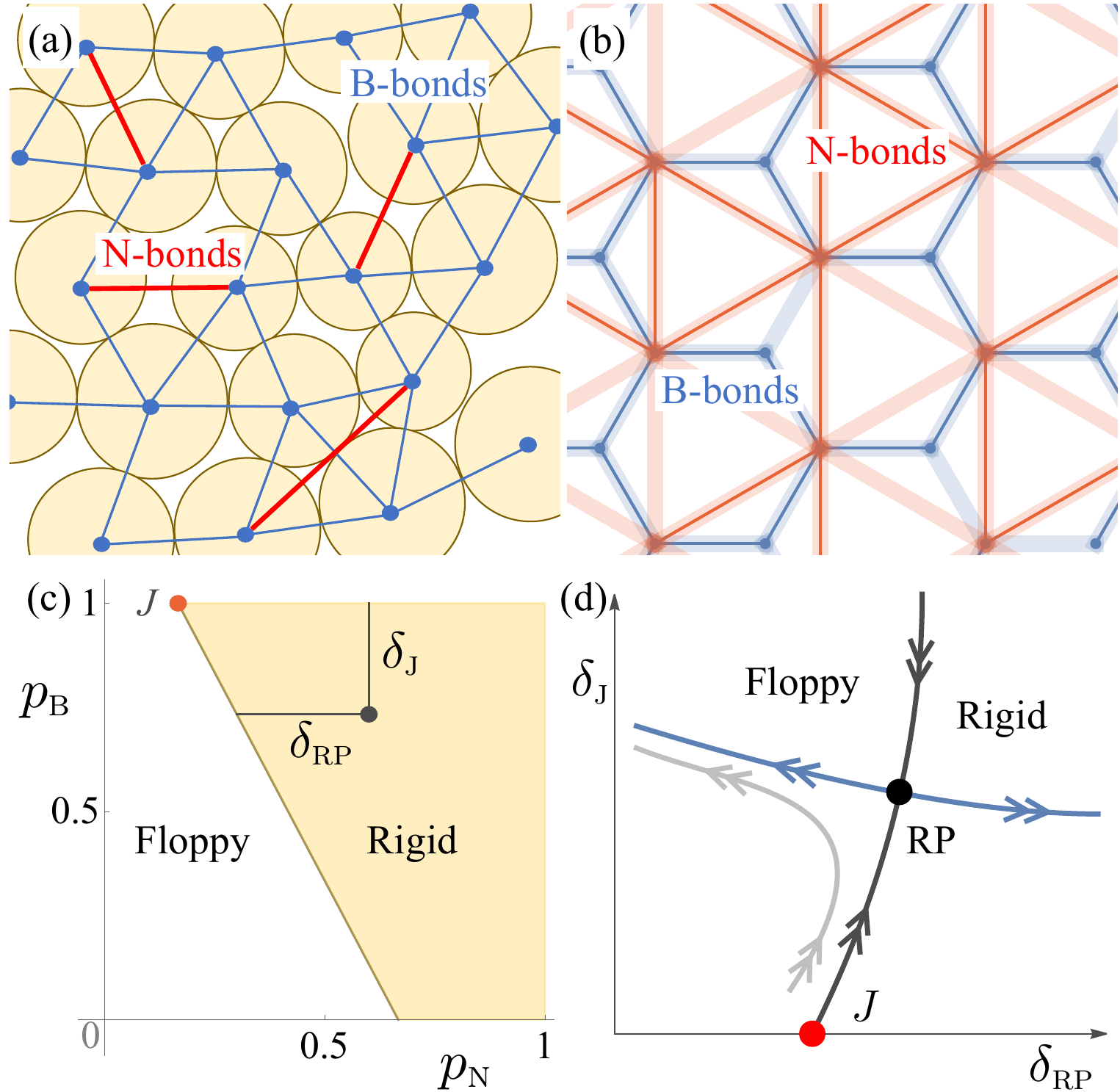} 
\end{center}
\caption{(a) Jammed disk packing, underlying contact network (B-bonds in
blue) and randomly added next-nearest neighbor N-bonds (red).
(b) HTL model with nearest and next-nearest neighbor bonds (solid blue and red
lines) connecting sites of a honeycomb lattice.
Faint blue and red lines show underlying honeycomb and triangular lattices,
respectively.
(c) Phase diagram of the HTL model in terms of occupation probabilities for B
and N-bonds.
The yellow region corresponds to the rigid state, and is separated from the floppy
state by an RP line ending at a jamming point $J$ (red disk).
(d) Conjecture for a crossover flow diagram projected into
$\DpRP \times \DpB$ space.
$J$ (red disk) and $RP$ (black disk) represent fixed points of a putative
renormalization-group scheme.
The blue, black and gray lines represent the unstable manifold, the critical line and a sample trajectory, respectively.
\label{fig:Model}}
\end{figure}

At jamming~\cite{LiuNag2010}, two-dimensional disk packings form a disordered contact
network [blue lines in Fig.~\ref{fig:Model}(a)] that supports compression but
not shear.
Mimicking compression by randomly adding next-nearest neighbor bonds between disks [red N-bonds
in Fig.~\ref{fig:Model}(a)] and/or randomly removing B-bonds can lead
to either jamming or RP depending on the population for each type of bond~\cite{LiarteLub2019}.
A simpler model that yields the same scaling behavior consists of randomly placing
`B' and `N'-bonds between nearest and next-nearest neighbor pairs of sites
[blue and red solid lines in Fig.~\ref{fig:Model}(b)] of a periodic honeycomb
lattice.
This network describes a diluted version of a 3-sub-lattice system consisting of
a honeycomb lattice [shaded blue in Fig.~\ref{fig:Model}(b)] and two triangular
lattices (shaded red; here we show only the bonds of one triangular lattice).
Detailed knowledge of the mechanical behavior of periodic lattices
allowed the development of an EMT at finite dimension%
~\footnote{See also Refs.~\cite{ZacconeSco2011,SchlegelZac2016} for calculations in finite dimension based on the nonaffine response of amorphous solids.}
for jamming~\cite{LiarteLub2019} and for the crossover from jamming to RP, valid in both rigid and floppy states.
We will employ these results to derive \emph{explicit solutions} for the
critical scaling of the susceptibilities of disordered viscoelastic matter near jamming and RP.
Our analysis not only allows for quick
assessment of scale-invariant behavior of quantities such as viscosities and
correlations (without the need for computationally-expensive simulations); it also serves as an example of how one may analyze rigidity transitions for which the universality class has not been determined.

Figure~\ref{fig:Model}(c) shows the phase diagram of the honeycomb-triangular
lattice (HTL) model in terms of occupation probability of nearest neighbor B bonds
and next-nearest neighbor N bonds.
Rigid (yellow) and floppy regions are separated by an RP line that terminates
in a multicritical jamming point $J$ (red disk).
From Fig.~\ref{fig:Model}(c), one can also extract definitions for the scaling
variables $\DpB$ and $\DpRP$, chosen so that $\DpRP=0$ at RP, and  $\DpB$ is also zero at jamming.

RP should generically be codimension one, because only one constraint
(isostaticity) needs to be satisfied. In the HTL model of Fig.~\ref{fig:Model}(b), jamming is codimension two.
But the jump in bulk modulus characteristic of jamming here demands a complete
honeycomb lattice; one can see that if the three orientations of hexagon bonds
were independently populated, the jamming transition would be codimension four
(their three probabilities set to one plus isostaticity).
This special tuning of the system to favor the bulk modulus is echoed in the
jamming of frictionless spheres, where the first state of self stress~\cite{LubenskySun2015} leads to
a jump in the bulk modulus because the conjugate degree of freedom (a uniform
compression) was used to tune the system to the rigidity transition.
As evidence for this, shear jamming of frictionless spheres has a jump in a
single anisotropic modulus~\cite{BaityJesiGLNS17}.

We conjecture that there is a class of disordered elastic systems for which a
renormalization-group scheme leads to the typical crossover flow
diagram~\cite{Cardy1996} (projected in $\DpRP \times \DpB$ space) illustrated
in Fig.~\ref{fig:Model}(d).
The scaling variable $\DpRP \propto \Delta z \equiv z-z_c$ must be relevant for
both jamming and RP, but the depletion probability of the B-lattice $\DpB$
is relevant only for jamming.
This behavior is captured by the direction of the arrows coming in and out of
the putative jamming and RP fixed points (red and black disks, respectively) in
Fig.~\ref{fig:Model}(d).
A system near the $J$ fixed point ($\DpB, |\DpRP| \ll 1$) will be
controlled either by $J$ if a crossover variable
$\DpB /|\DpRP|^\varphi \ll 1$ for some exponent $\varphi$, or by RP if
$\DpB /|\DpRP|^\varphi \gg 1$, i.e. for trajectories such as the gray line passing sufficiently close to the critical line (black solid line.)
Though $\DpB$ does not have a direct interpretation in the jamming of sphere
packings [except for the network model of Fig.~\ref{fig:Model}(a)], there might be
variables that play a similar role, such as attractive interactions in soft
gels~\cite{GadoMao2020}.

We now introduce a scaling \emph{ansatz} for the longitudinal response
function~\cite{ChaikinLub1995} near jamming:
\begin{equation}
    \frac{\XL}{\chi_0}
        \approx |\DpRP|^{-\gamma} \mathcal{L} \left( \frac{q/q_0}{|\DpRP|^{\nu}},
            \frac{\omega/\omega_0}{|\DpRP|^{z \nu}},
            \frac{\DpB/\delta_0}{|\DpRP|^{\varphi}}\right),
\label{eq:ResponseScaling}
\end{equation}
where $q$ is the wavevector, $\omega$ is the frequency, $\gamma$, $\nu$, $z$ and
$\varphi$ are critical exponents for the susceptibility, correlation length,
correlation time, and crossover behavior, respectively~\cite{Cardy1996,Sethna2006},
and $\mathcal{L}$ is a universal scaling function.
The constants $\chi_0$, $q_0$, $\omega_0$ and $\delta_0$ are nonuniversal scaling factors.
Many other properties can be derived from $\mathcal{L}$ (Table~\ref{tab:Functions}).
Such space-time susceptibilities, and the corresponding structure and
correlation functions, are the fundamental linear response quantities for
materials.
They have been well studied in glassy systems, but have hitherto not been a focus
in the study of jamming or RP.
Baumgarten et al.~\cite{BaumgartenVT17} and Hexner et al.~\cite{HexnerNLPoster}
have studied the static response of frictionless jammed spheres to a sinusoidal
perturbation; they find diverging length scales that are different from the ones presented here.
Because our system is on a regular lattice, and particularly because our analysis
replaces the disordered lattice with a uniform one, it is natural for us to fill
this gap.

Our approach goes beyond previous work~\cite{GoodrichSet2016} in
two aspects.
First, rather than starting with an ansatz for the free energy in terms of
the excess contact number $\Delta z$, excess packing fraction $\Delta \phi$, shear
stress $\epsilon$ and system size $N$, we consider the longitudinal response
in terms of $\DpRP$, $q$, $\omega$ and $\DpB$.
Our variable $\DpRP$ is proportional to $\Delta z$.
Though we do not consider an explicit dependence of $\XL$ on $\epsilon$ or
$\Delta \phi$%
~\footnote{Note that $\DpRP$ does not change with lattice deformation for our
system.
This contrasts with the case of compressed disks in which $\Delta z$ can vary
with $\Delta \phi$.
We assume fixed (quenched) disorder in our model.}%
, we can extract equivalent expressions for moduli and correlations from the
dependence of $\XL$ on $q$.
Importantly, the inclusion of $\omega$ in our analysis allows us to predict
dynamical properties such as viscosities.

Second, we use EMT~\cite{LiarteLub2019} to derive and validate both the universal exponents and the universal scaling functions ($\mathcal{L}$), for both jamming and RP.
This form of EMT is based on the coherent-potential
approximation~\cite{ElliottLea1974,FengGar1985} (CPA), and is known to reproduce
well results obtained from simulations of randomly-diluted lattices with
two-body%
~\footnote{
A more sophisticated version of EMT is needed to reproduce the scaling
behavior of randomly-diluted lattices with three-body forces such as
bending~\cite{LiarteLub2016}.
}
harmonic interactions~\cite{SchwartzSen1985,LiarteLub2020}, even for
undamped~\cite{MaoLub2011,LiarteLub2019} and overdamped
dynamics~\cite{YuchtBro2013,DueringWya2013}.
Although the CPA involves mean-field-like uncontrolled approximations, it
preserves the topology of the original lattices --- an essential ingredient that
ultimately allows one to describe jamming.
Here we focus on the longitudinal response, since the full response of isotropic elastic
systems can be decomposed into longitudinal and transverse
components, and the latter has the same scaling form near both jamming and RP as
the longitudinal response near RP; see~\cite{LiarteSet2022}.

We use the long wavelength limit of the longitudinal response
$\XL$ along with EMT results from
Ref.~\cite{LiarteLub2019} to derive critical exponents (see
Table~\ref{tab:Exponents}) and the universal scaling function $\mathcal{L}$ in
Eq.~\eqref{eq:ResponseScaling} (see~\cite{SM21,LiarteSet2022}),
\begin{equation}
    \mathcal{L}( u, v, w )
        = \left[\frac{u^2}{1+w / \left(\sqrt{1 - \tilde{v}(v)} \pm 1\right)}
            - \tilde{v} (v)\right]^{-1},
    \label{eq:Ldef}
\end{equation}
where $\tilde{v} (v) = v^2$ and $i \, v$ for undamped and overdamped dynamics, respectively, and the plus and minus signs correspond to solutions in
the elastic and floppy states, respectively. 
Equation~\eqref{eq:Ldef} embodies the central results of this paper.
From Eqs.~\eqref{eq:ResponseScaling} and~\eqref{eq:Ldef}, we will extract the
universal behavior of the elastic moduli, viscosities as well as the density
response and correlation functions (dynamic structure factor).
Though it is not certain that these functions are as universal as critical
exponents, recent simulations of compressed hyper-spheres~\cite{SartorCor2021}
indicate that critical amplitudes calculated using mean-field models at infinite
dimension are preserved for low-dimensional jammed packings.
\begin{table}[h!]
\centering
\begin{tabular}{ |l|c|c|c|c|c|c| }
\hline
 & $\gamma$ & $z$ & $\nu$ & $\varphi$ & $\beta_\textsc{b}$ & $\gamma_\textsc{b}$ \\
 \hline
 Jamming & 2 & 1 (2) & 1 & 1 & 0 & 1 (2)  \\
 Rigidity Percolation &  2  & 2 (4) & 1/2 & - & 1 & 0 (1)  \\
\hline
\end{tabular}
\caption{\label{tab:Exponents} Critical exponents for the longitudinal
susceptibility ($\gamma$), correlation length ($\nu$), correlation time ($z$)
and crossover behavior ($\varphi$) near jamming and RP for undamped and overdamped
(between parentheses if different from undamped) dynamics.
The exponents $\beta_\textsc{b}$ and $\gamma_\textsc{b}$ can be derived from $\gamma$, $\nu$ and
$z$ (see Table~\ref{tab:Functions}), and describe power-law singularities for the
bulk modulus and viscosity, respectively.}
\end{table}

For $|\DpRP| \ll \DpB$ [$w\gg 1$ in Eq.~\eqref{eq:Ldef}], our model
exhibits RP criticality: $\DpB$ becomes an irrelevant variable, and
$\mathcal{L}(u,v,w) \rightarrow \bar{\mathcal{L}}(u,v)$, with
\begin{equation}
    \bar{\mathcal{L}}(u,v)
        = \left[u^2 \left(\sqrt{1 - \tilde{v}(v)} \pm 1\right)
            - \tilde{v} (v)\right]^{-1}.
    \label{eq:LbarDef}
\end{equation}
Here the change in $\mathcal{L}$ is accompanied by a change in the critical exponents $\nu$ and $z$ (see 
Table~\ref{tab:Exponents}).
Note that the exponent $z \nu$ depends only on the type of dynamics, but the exponent $\nu$ equals $1$ and $1/2$ for jamming and RP, respectively.

Our formulation of Eqs.~(\ref{eq:ResponseScaling}-\ref{eq:LbarDef}) represents a deliberate effort to emphasize model-independent \emph{(universal)} features.
Note e.g. that our model definition of the non-universal scaling factor $q_0$ is different for jamming and RP; the latter involves a term that increases as one moves away from the jamming multicritical point.
Besides, our formulation allows for the suitable incorporation of analytic corrections to scaling~\cite{AharonyFis1980,AharonyFis1983,Cardy1996,RajuSet2019}, which can be added in a case-by-case basis.
In general, we expect these corrections to appear through the introduction of nonlinear scaling fields, $u_q (q, \omega, \delta_\textsc{j}) = q/q_0 + \dots$, $u_\omega (q, \omega, \delta_\textsc{j}) = \omega/\omega_0 + \dots$, $u_\textsc{j} (q, \omega, \delta_\textsc{j}) = \delta_\textsc{j}/\delta_{0} + \dots$, which would replace $q/q_0$, $\omega / \omega_0$ and $\delta_\textsc{j} / \delta_0$ in Eq.~\eqref{eq:ResponseScaling}. Here the dots represent higher-order terms and perhaps linear terms in the other variables (rotating the axes). These nonlinear scaling fields can be viewed as the difference between the lab parameters and Nature's natural variables, or as the coordinate transformation removing the (hypothetical) nonlinear terms in the renormalization group to their hyperbolic normal form~\cite{RajuSet2019}.
In order to use our scaling predictions to describe behavior far from the critical point, one must first determine the appropriate scaling fields $u_q$, $u_\omega$ and $u_\textsc{j}$ for the particular system.

\begin{figure}[ht]
\begin{center}
\includegraphics[width=\linewidth]{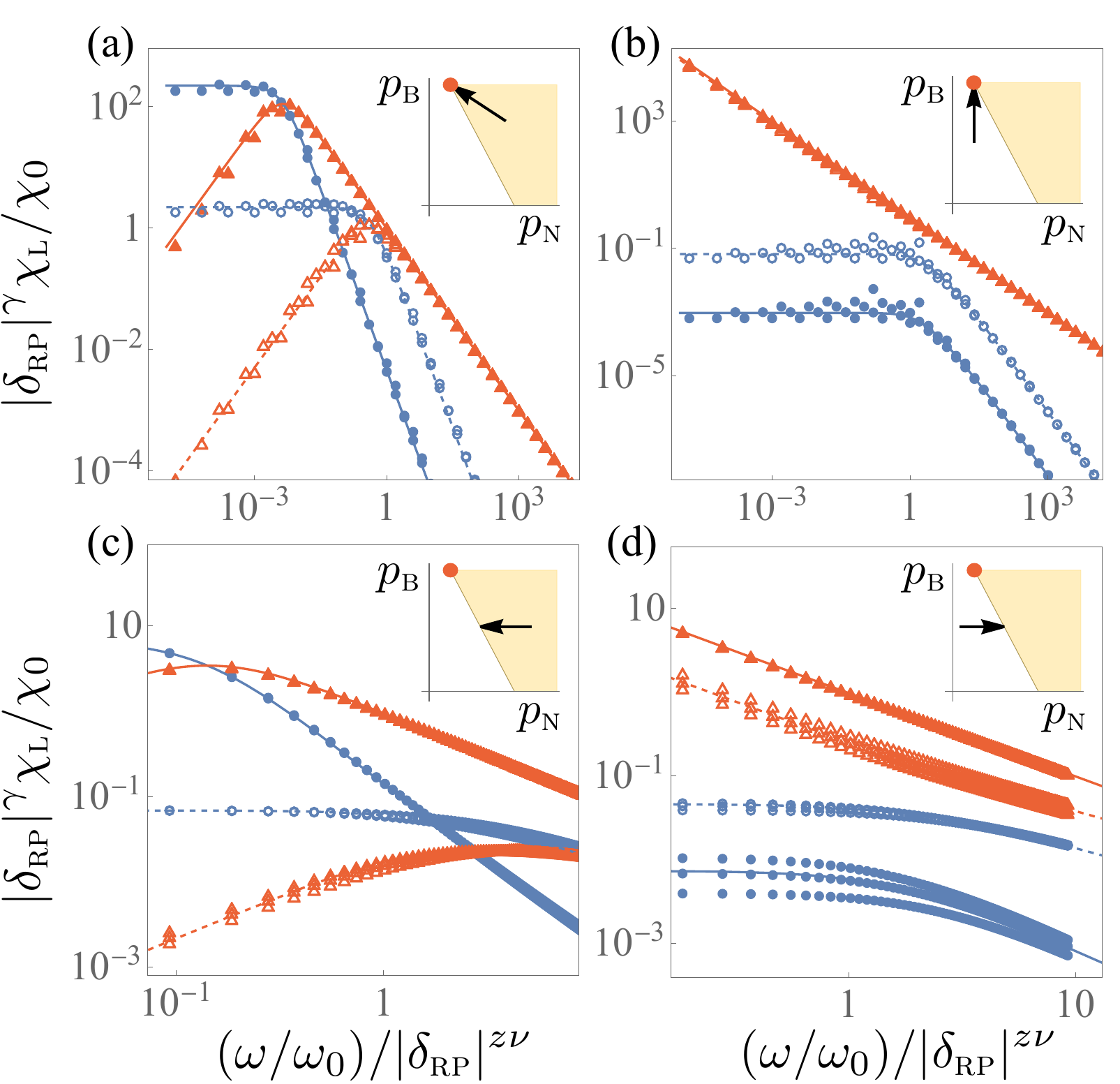} 
\end{center}
\caption{
Scaling collapse plots showing the universal behavior of the longitudinal response as a function of rescaled frequency near jamming (first row) and RP (second row), for \emph{overdamped} dynamics.
Blue disks and red triangles are full solutions of the EMT equations for the real and imaginary parts of $|\DpRP|^\gamma \XL / \chi_0$, respectively.
Solid and dashed curves are the universal scaling predictions of Eqs.~\eqref{eq:Ldef} and~\eqref{eq:LbarDef}.
We consider points approaching jamming and RP along the paths indicated in the inset graphs of each panel.
We use $q/|\DpRP|^{\nu}=0.1$ (closed symbols) and $1$ (open symbols) in all panels, and $\DpB/|\DpRP|^\varphi$ equal to $\sqrt{5}/4$ from the rigid side~(a), and equal to $2$ from the floppy side~(b).
Full solutions run at $|\DpRP| = 10^{-2}$, $10^{-3}$, and $10^{-4}$ for RP and a range $|\DpRP|\in [5\times10^{-2}, 5\times10^{-6}]$ for jamming show convergence to our universal asymptotic predictions.
\label{fig:CollapseOD}}
\end{figure}

Equation~\eqref{eq:ResponseScaling} implies that solutions for $|\DpRP|^\gamma \XL$ as a function of one of the three invariant scaling combinations (the other two kept constant) should lie on the curves given by Eqs.~\eqref{eq:Ldef} and~\eqref{eq:LbarDef}, respectively.
Hence, plots for different values of $|\DpRP|$ should collapse for several paths approaching jamming or RP.
Figure~\ref{fig:CollapseOD} shows an example of a scaling collapse plot of the rescaled longitudinal response as a function of rescaled frequency for overdamped dynamics at fixed $q/|\DpRP|^\nu$ and $\delta_\textsc{j}/|\DpRP|^\varphi$, and for paths approaching jamming (first row) and RP (second row) from both the rigid and floppy phases (see inset in each panel).
Real parts are in blue; imaginary (dissipative) parts in red.
The solid and dashed curves are the asymptotic universal scaling predictions  [Eqs.~\eqref{eq:Ldef} and~\eqref{eq:LbarDef}] at two different values of the wavevector scaling variable $q/|\DpRP|^\nu$.
Although there are model-specific predictions for the nonuniversal scaling factors, we choose them to best fit the collapsed data.

The collapses of Fig.~\ref{fig:CollapseOD} not only validate our universal scaling forms; they indicate an interesting crossover to a regime dominated by dissipation (the imaginary part of $\XL$ in red) as the frequency increases.
Note that the real part $\mathcal{L}^\prime (v)$ plateaus and the imaginary part $\mathcal{L}^{\prime \prime} (v)$ (the dissipation) vanishes at low frequency $v$.
At high frequency, both $\mathcal{L}^{\prime}$ and $\mathcal{L}^{\prime \prime}$ decay to zero, but $\mathcal{L}^{\prime}$ decays faster than $\mathcal{L}^{\prime \prime}$, except in the limit of very large $u$ and $v$, where both $\mathcal{L}^{\prime}$ and $\mathcal{L}^{\prime \prime}$ decay as $v^{-1/2}$.
Hence, there is a frequency $\omega$ in which $\mathcal{L}^\prime \sim \mathcal{L}^{\prime \prime}$, and above which the response is dominated by the dissipative imaginary part.
From Eq.~\eqref{eq:Ldef}, we find that $\omega \sim D^* q^2$ in this regime, leading to the definition of an effective diffusion constant $D^*\sim |\DpRP|^{(z-2)\nu}$.
Using the exponents shown in Table~\ref{tab:Exponents}, we find that $D^* \sim \mathcal{O}(1)$ and $\sim |\DpRP|$ for jamming and RP, respectively.
In terms of rescaled variables, this crossover happens at $v\sim u^2$ for both transitions.
In the liquid phase [(b) and (d)], $\mathcal{L}^\prime$ behaves as in the elastic phase, but $\mathcal{L}^{\prime \prime}$ diverges rather than vanishing at low $v$ due to the predominant viscous response of the fluid state.

\begin{figure*}[!ht]
\begin{center}
\includegraphics[width=\linewidth]{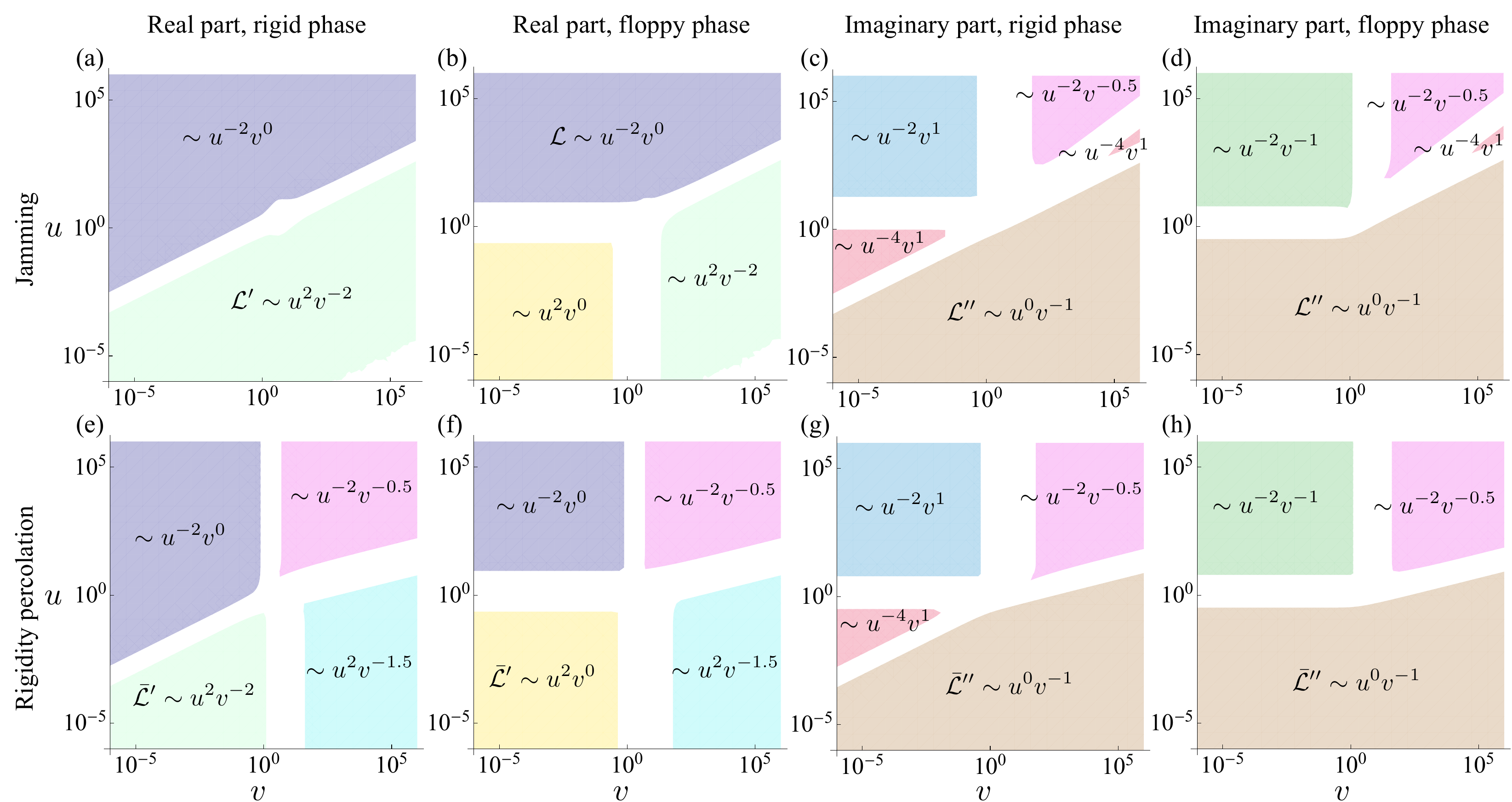}
\end{center}
\caption{{\bf Overdamped asymptotic exponents for universal longitudinal response.} Diagram in the $u$ (rescaled wavevector) $\times v$ (rescaled frequency) plane, showing regions of distinct power-law behavior of the jamming (first row) and RP (second row) universal scaling functions for overdamped dynamics in both the rigid and floppy phases.
The first and second (third and fourth) columns correspond to the real (imaginary) parts of $\mathcal{L}$ and $\bar{\mathcal{L}}$.
We use $w=1$ for jamming.
\label{fig:GlobalPlotsOD}}
\end{figure*}

Equations~(\ref{eq:Ldef}) and (\ref{eq:LbarDef}) also imply that our universal functions for the longitudinal response $\mathcal{L}(u,v,w)$ and $\bar{\mathcal{L}}(u,v)$ generally behave as $u^\alpha v^\beta$ with the exponents $\alpha$ and $\beta$ depending on the region in the $u \textrm{ (rescaled wavector)} \times v \textrm{ (rescaled frequency)}$ plane.
To illustrate and map this global behavior, we show in Fig.~\ref{fig:GlobalPlotsOD} the power-law regions for which $\mathcal{L}(u,v,w) \propto u^\alpha v^\beta$ and $\bar{\mathcal{L}}(u,v) \propto u^\alpha v^\beta$, with $(\alpha, \beta)$ very close to their asymptotic values.
The first and second rows correspond to our scaling forms for jamming and RP, respectively.
To generate each panel, we numerically calculate the exponents using $f_\alpha \equiv \partial \log \mathcal{L}/\partial \log u$ and $f_\beta \equiv \partial \log \mathcal{L}/\partial \log v$ for jamming and similar formulas for RP.
We then plot the regions in which $|f_\alpha - \alpha| < 0.1$ and $|f_\beta - \beta| < 0.1$, for several values of $\alpha$ and $\beta$.

Figure~\ref{fig:GlobalPlotsOD} offers a vivid pictorial view allowing an easier assessment of the global behavior associated with our universal forms for jamming and rigid percolation.
By comparing the two rows, notice how the change in universality class is also reflected in the behavior of the universal scaling functions.
For instance, although jamming and RP exhibit similar qualitative features for the imaginary part [(c), (d), (g) and (h)], RP shows additional regimes for the real part, which do not appear in jamming [compare e.g. (a) and (e) or (b) and (f)].

In the Supplementary Material~\cite{SM21}, we present results for undamped dynamics that are analogous to Figs.~\ref{fig:CollapseOD} and~\ref{fig:GlobalPlotsOD} in this letter. 
The full solutions of our effective-medium theory equations also converge to our universal scaling functions, except in the limit of very low frequencies.
In fact, the asymptotic solutions derived in~\cite{LiarteLub2019} do not capture the small but nonzero imaginary parts of the effective spring constants at frequencies smaller than $\sim \omega^*$ (the characteristic crossover to isostaticity) when there is no damping.
This feature has important consequences for energy dissipation in systems believed to exhibit behavior related to RP.
The corrections to scaling appear as singular perturbations to the self-consistency equations and vanish as powers of $|\DpRP|$ in dimensions larger than three.
Moreover, the scaling variables contain logarithms in two dimensions.
This analysis is beyond the scope of the present work, and will be presented in a separate manuscript.

Equations~\eqref{eq:ResponseScaling} and~\eqref{eq:Ldef} determine the scaling
behavior of several quantities characterized by the general form,
\begin{equation}
    \frac{Y}{Y_0}
        = |\DpRP|^{y} \, \mathcal{Y} \left( \frac{q/q_0}{|\DpRP|^{\nu}},
            \frac{\omega/\omega_0}{|\DpRP|^{z \nu}},
            \frac{\DpB/\delta_0}{|\DpRP|^{\varphi}}\right),
    \label{eq:Ydef}
\end{equation}
where in Table~\ref{tab:Functions} we present explicit expressions for the exponent $y$ and universal
function $\mathcal{Y}$ describing the bulk modulus ($B$), viscosity ($\zeta$),
density response ($\Pi$) and correlation function ($S$).
The behavior near RP is obtained by replacing $\mathcal{Y}$ and $\mathcal{L}$
in the third column of Table~\ref{tab:Functions} by $\bar{\mathcal{Y}}$ and
$\bar{\mathcal{L}}$ (now functions of $u$ and $v$
only), respectively, along with appropriate changes for the exponents (see
Table~\ref{tab:Exponents}).
The scaling behavior of the shear modulus and viscosity near jamming and RP
is the same as that of $B$ and $\zeta$, respectively, near RP.
\begin{table}[!ht]
\centering
\begin{tabular}{ |c|c|c| }
\hline
$Y$ & $y$ & $\mathcal{Y}$  \\
 \hline
$B$ & $\beta_\textsc{b} \equiv \gamma -2 \nu$ &
    $\mathcal{B} = (\partial \mathcal{L}^{-1} / \partial u ) / (2\,u)$  \\
$\zeta$ & $-\gamma_\textsc{b} \equiv \gamma-(2+z)\nu$  &
    $\mathcal{Z}
        = (1/v)\, \mathrm{Im} \left[\mathcal{B} \right]$ \\
$\Pi$ & $2\nu-\gamma$  &
    $\mathcal{P}
        = u^2 \mathcal{L}$ \\
$S$ & $(2+z)\nu - \gamma$  &
    $\mathcal{S}
        = (1/ v)\, \mathrm{Im} [\mathcal{P}]$ \\
\hline
\end{tabular}
\caption{Critical exponent $y$ and universal scaling function $\mathcal{Y}$
describing the singular behavior of the bulk modulus $B$ and viscosity $\zeta$,
density response $\Pi$ and correlation function $S$, according to
Eq.~\eqref{eq:Ydef}.
\label{tab:Functions}}
\end{table}

\begin{figure}[!ht]
\begin{center}
\includegraphics[width=\linewidth]{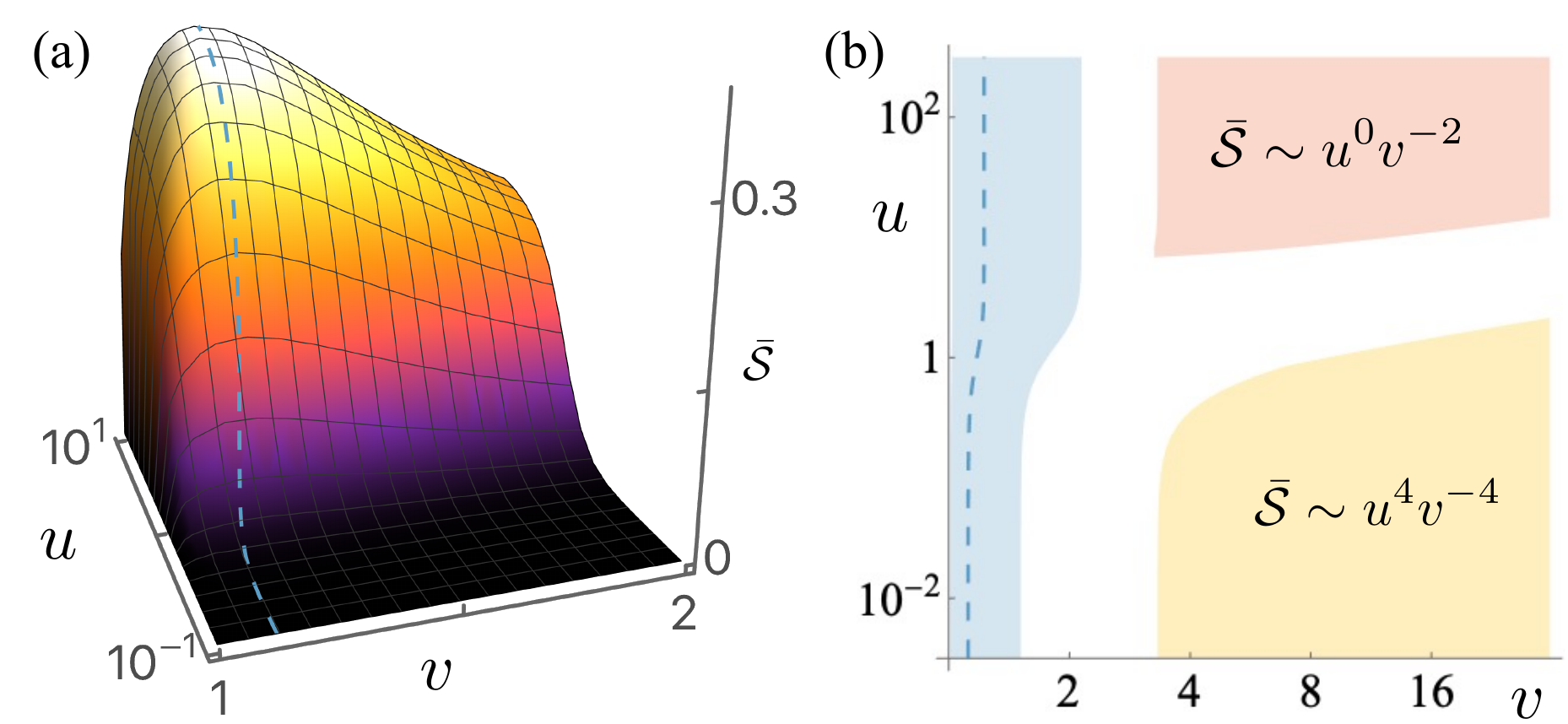} 
\end{center}
\caption{(a) 3D plot of the universal scaling function for the correlation function
$\bar{\mathcal{S}} (u, v)$, for undamped fluids near RP.
The blue dashed line corresponds to the rescaled frequency $v^*$ (the boson peak)
at which $\bar{\mathcal{S}} (u,v)$ is maximum for fixed rescaled wavevector $u$.
(b) $u\times v$ diagram showing the boson peak (blue dashed line) and power law regions for
which $\mathcal{S} (u, v) \propto u^{\alpha}v^{\beta}$, with $(\alpha,\beta)$ close to their asymptotic values $(0,-2)$ (red)
and $(4,-4)$ (yellow).
In the blue region the condition $\bar{\mathcal{S}}(u,v)>\bar{\mathcal{S}}(u,v^*)/2$
is satisfied.
\label{fig:UrsellUDL}}
\end{figure}

To illustrate the broad applicability of our scaling forms, we discuss our results for the density-density correlation --- the structure function for isotropic fluids at $q\neq 0$.
Figure~\ref{fig:UrsellUDL}(a) shows a 3D plot of the universal function $\bar{\mathcal{S}}(u,v)$ (see Table~\ref{tab:Functions}) for undamped fluids near RP.
At fixed $u$, $\bar{\mathcal{S}} (u, v)$ has a maximum (blue dashed line) at
$v=v^* \approx \mathcal{O} (1)$ [i.e. $\omega^* \propto \DpRP$] (see~\cite{LiarteSet2022}), which coincides with the crossover from Debye to isostatic behavior, interpreted as the paradigmatic \emph{boson peak}~\cite{VitelliNag2010,KoehlerSch2013,GiuliWya2014,FranzZam2015} of glasses~\cite{BinderKob2011}.
Near jamming or RP, this point marks the onset of the enhancement of the population
of low-energy modes~\cite{SilbertNag2005} leading to a flat density of states at
low frequency~\cite{SilbertNag2005,LiarteLub2019}.
At fixed $v$, $\bar{\mathcal{S}}$ plateaus at a value of $u$ of $\mathcal{O}(1)$
(i.e. at $q \propto |\DpRP|^{1/2}$).
Our explicit formulas also provide a simple tool to map the global behavior of many quantities of interest.
For example, Fig.~\ref{fig:UrsellUDL}(b) shows a diagram in terms of rescaled wavevector $u$ and frequency $v$ marking the boson peak (blue-dashed line) and regions where $\bar{\mathcal{S}} (u, v)$ exhibits power-law behavior.
The blue region indicates the neighborhood of the boson peak, in which
$\bar{\mathcal{S}}(u,v) > \bar{\mathcal{S}}(u,v^*)/2$, and the red and yellow regions show power-law regimes in $u$ and $v$.

Near jamming, the two-time density-density correlation function $S_{nn}(r- r^\prime, t- t^\prime)$ in real space is given by,
\begin{align}
    & S_{nn}(r, r^\prime, t, t^\prime)/S_0
        \nonumber \\
    & \approx |\DpRP|^{(2+D)\nu-\gamma}
        \mathcal{S}
            \left(\frac{(r-r^\prime)/\ell_0}{|\DpRP|^{-\nu}}, \frac{(t-t^\prime) / t_0}{|\DpRP|^{-z\nu}},
            \frac{\DpB/\delta_0}{|\DpRP|^\varphi}\right),
    \label{eq:C_Scaling}
\end{align}
where $\ell_0$ and $t_0$ are nonuniversal scaling factors, and
\begin{equation}
    \mathcal{S}(\rho,s,w)
        = \displaystyle\int d\bm{u} \, dv \, e^{i (u \cdot \rho - v s)} \,
            \frac{\mathrm{Im} \, \mathcal{P}(u,v,w)}{v},
\end{equation}
where $\rho$ and $s$ are arguments for the universal scaling function $\mathcal{S}$ associated with the re-scaled distance and time, respectively.
The behavior near RP is obtained by replacing $\mathcal{S}$ and $\mathcal{P}$ by $\bar{\mathcal{S}}$ and $\bar{\mathcal{P}}$, respectively, along with appropriate changes for the exponents (see Table~\ref{tab:Exponents}).
Equation~\eqref{eq:C_Scaling} and the corresponding equation for RP lead to definitions of diverging length and time scales, $\ell = |\DpRP|^{-\nu} \ell_0$ and $\tau = |\DpRP|^{-z\nu} t_0$, respectively.
Our characteristic length scale diverges as $|\DpRP|^{-1}$ for jamming, and as $|\DpRP|^{-1/2}$ for RP.
These divergences should be compared with traditional definitions of $\ell_c \sim |\Delta z|^{-1/2}$ and $\ell^* \sim |\Delta z|^{-1}$, as discussed in the literature~\cite{EllenbroekSaa2006,LernerWya2014,KarimiMal2015,BaumgartenVT17,HexnerNLPoster},
Note that $\nu$ in our paper should not be confused with exponents for the finite-size scaling of the probability density $\Delta \sim L^{1/\nu}$, as reported in~\cite{JacobsTho1995}.

In this letter, we have combined scaling theory and the EMT of
Ref.~\cite{LiarteLub2019} to produce analytical formulas for universal scaling
functions for the longitudinal dynamical response near both jamming and RP.
Our equations can be used to determine the space-time dependence of universal
functions for several quantities (such as moduli, viscosities and correlations)
near the onset of rigidity in both the solid and liquid phases.
A direct approach to experimentally validate our predictions consists of using
3D printers to fabricate and perform experiments on the disordered elastic
networks illustrated in Figs.~\ref{fig:Model}(a) and (b). We also expect these scaling forms to apply to more traditional glass forming systems such as colloidal suspensions. Here, in addition to more standard scattering measurements, new techniques for measuring 3D particle positions and even stresses with high precision may make it feasible to measure these functional forms and test our predictions \cite{weeks2000three,lin2016measuring,bierbaum2017light,leahy2018quantitative}.
In such suspensions, we expect that the scaling functions will capture the behavior in the elastic regime. However, our
theory is built on a fixed network topology and lacks some features of the liquid
phase.
Annealed rather than quenched disorder~\cite{Cardy1996} (or even intermediate
disorder~\cite{CarmoSal2010}) could be needed to describe
viscoelastic fluids.
An extension of our analysis includes an investigation~\cite{ThorntonCho2022} of the intriguing connections between the featureless low-energy modes in our system and the unconventional \emph{particle-hole continuum} measured using momentum and energy-resolved spectroscopic probes in certain \emph{strange metals}~\cite{MitranoAbb2018,HusainAbb2019}.
Other extensions could include the incorporation of \emph{anisotropic}
bond occupation~\cite{ZhangDas2014}, which plays a major role in the crossover scaling of thickening suspensions near frictional jamming~\cite{RamaswamyCoh2021} and that can lead to simpler models for both
shear jamming~\cite{BehringerCha2018} and thickening~\cite{BrownJae2014}, as well as the incorporation of random stress fields, which can elucidate the unjamming of colloidal suspensions (such as titanium dioxide) due to activity~\cite{HenkesMar2011}.

\begin{acknowledgments}
We thank Andrea Liu, Bulbul Chakraborty, Daniel Hexner, Eleni Katifori, Emanuela del Gado, Itay Griniasty, Matthieu Wyart, Meera Ramaswamy, Peter Abbamonte, Sean Ridout, Tom Lubensky and Xiaoming Mao for useful conversations.
This work was supported in part by NSF DMR-1719490 (SJT and JPS), NSF CBET Award \verb|#| 2010118 (DBL, ES, JPS, and IC) and NSF CBET Award \verb|#| 1509308 (ES and IC).
DBL also thanks financial support through FAPESP grants \verb|#| 2016/01343-7 and \verb|#| 2021/14285-3.
DC is supported by a faculty startup grant at Cornell University.
\end{acknowledgments}

%

\end{document}